\documentclass[runningheads]{llncs}

\usepackage[T1]{fontenc}
\usepackage{graphicx}
\usepackage{amsmath}
\usepackage{url}
\usepackage{booktabs}

\newcommand{\placeholderfigure}[1]{%
\fbox{\begin{minipage}[c][0.24\textheight][c]{0.92\textwidth}
\centering Figure placeholder\\[0.5em]
#1
\end{minipage}}%
}

\newcommand{\figurefile}[2]{%
\IfFileExists{#1}{%
\includegraphics[width=\textwidth]{#1}%
}{%
\placeholderfigure{#2\\[0.5em]\texttt{#1}}%
}%
}

\begin{document}

\title{A High-Fidelity Cardiac Digital Twin for Assessing Rate-Dependent Responses: A Comparative Multi-scale Study of TT2 and Tomek Models in Stratified 3D Ventricles}

\titlerunning{A High-Fidelity Cardiac Digital Twin}

\author{Zhenyin Fu\inst{1} \and Zhaokai Kong\inst{1} \and Lucheng Xu\inst{2} \and Mingfeng Jiang\inst{3} \and Ling Xia\inst{1} \and Wenlong Xu\inst{4} \and Ruiqing Dong\inst{5} \and Dongdong Deng$^\dagger$\inst{2}}

\authorrunning{Z. Fu et al.}

\institute{Zhejiang University, Hangzhou, China
\and
Dalian University of Technology, Dalian, China\\
\and
Zhejiang Sci-Tech University, Hangzhou, China
\and
China Jiliang University, Hangzhou, China
\and
The Fourth Affiliated Hospital of Soochow University, Suzhou, China\\\email{$^\dagger$dengdongdong@dlut.edu.cn}}

\maketitle

\begin{abstract}
Cardiac digital twins (CDTs) provide a computational platform for personalized cardiac modeling, but their predictive behavior depends strongly on the selected ionic model and the representation of transmural electrophysiological heterogeneity.  In this study, we developed a multiscale CDT validation framework that integrates physiologically optimized hierarchical configurations. Using single-cell (0D) models and three-dimensional (3D) patient-specific ventricular models, we implemented the ten Tusscher-Panfilov model (TT2) and Tomek ionic model and compared their electrophysiological responses across a range of pacing frequencies, with basic cycle lengths (BCLs) from 600 to 4000 ms. The results show that tissue-level electrical coupling substantially reshapes the intrinsic dynamical properties observed at the single-cell level. By benchmarking model outputs against clinical data, this study quantifies fidelity differences between the two ionic models across multiple scales and provides a scientific basis for ionic model selection and CDT construction.

\end{abstract}

\section{Introduction}

Cardiac Digital Twins (CDTs) are emerging as transformative computational platforms for precision cardiovascular medicine ~\cite{deng2019characterizing,tong2021preliminary,miao2026theoretical}. By integrating patient-specific anatomical imaging with biophysical simulation, CDTs support individualized decision-making in arrhythmia risk stratification, drug screening, and therapy optimization. However, the predictive fidelity of a CDT is strictly constrained by its representation of ventricular electrophysiological heterogeneity ~\cite{maharaj2008role}. The ventricular wall comprises functionally stratified layers, the epicardium (Epi), mid-myocardium (M-cell), and endocardium (Endo), each possessing distinct electrophysiological properties. Neglecting this transmural heterogeneity prevents CDTs from accurately simulating repolarization dispersion and the resulting substrates for reentrant arrhythmias.

Experimental evidence indicates that transmural APD relationships are strongly rate-dependent~\cite{antzelevitch2004electrophysiological}. At long BCLs, transmural APD dispersion may be amplified in specific experimental preparations~\cite{akar2003transmural}, potentially increasing repolarization heterogeneity and susceptibility to reentrant activity~\cite{glukhov2010transmural}. However, whether different human ventricular ionic models reproduce this trend, and whether it persists after tissue-level electrotonic coupling, remains unclear. 

Another central challenge lies in the selection of the ionic model engine. Preeminent human ventricular models, such as the ten Tusscher-Panfilov (TT2) ~\cite{ten2004model,ten2006alternans} and Tomek ~\cite{tomek2019development} models, possess divergent mathematical formulations for key ion channel kinetics, leading to distinct rate-adaptation characteristics~\cite{coppini2013late}. During the process of model translation, the BCL-sensitivity of these single-cell engines directly dictates the system's ability to replicate rate-dependent phenomena. If a model fails to accurately simulate the APD plateau or the evolution of transmural gradients at large BCLs, the CDT's predictive fidelity is compromised. Thus, a CDT should be conceptualized as an engineered multi-scale system where the ionic engine and physical tissue architecture jointly determine predictive behavior.

Accordingly, this study established a multiscale comparative CDT framework spanning single-cell and three-dimensional ventricular scales. The TT2 and Tomek models were implemented within the same ventricular geometry using a fixed, previously optimized transmural stratification. The study had three objectives: (1) to characterize the intrinsic rate-dependent APD responses of the two ionic models at the single-cell scale; (2) to determine how three-dimensional electrotonic coupling modifies their model-specific transmural gradients; and (3) to qualitatively compare the simulated rate-dependent trends with published canine experimental observations~\cite{akar2003transmural,kong2009transmural}. This framework was designed to clarify how ionic-model selection and tissue-level integration jointly influence rate-dependent CDT behavior.

\section{Materials and Methods}

\begin{figure}[htbp]
\centering
\includegraphics[width=\linewidth]{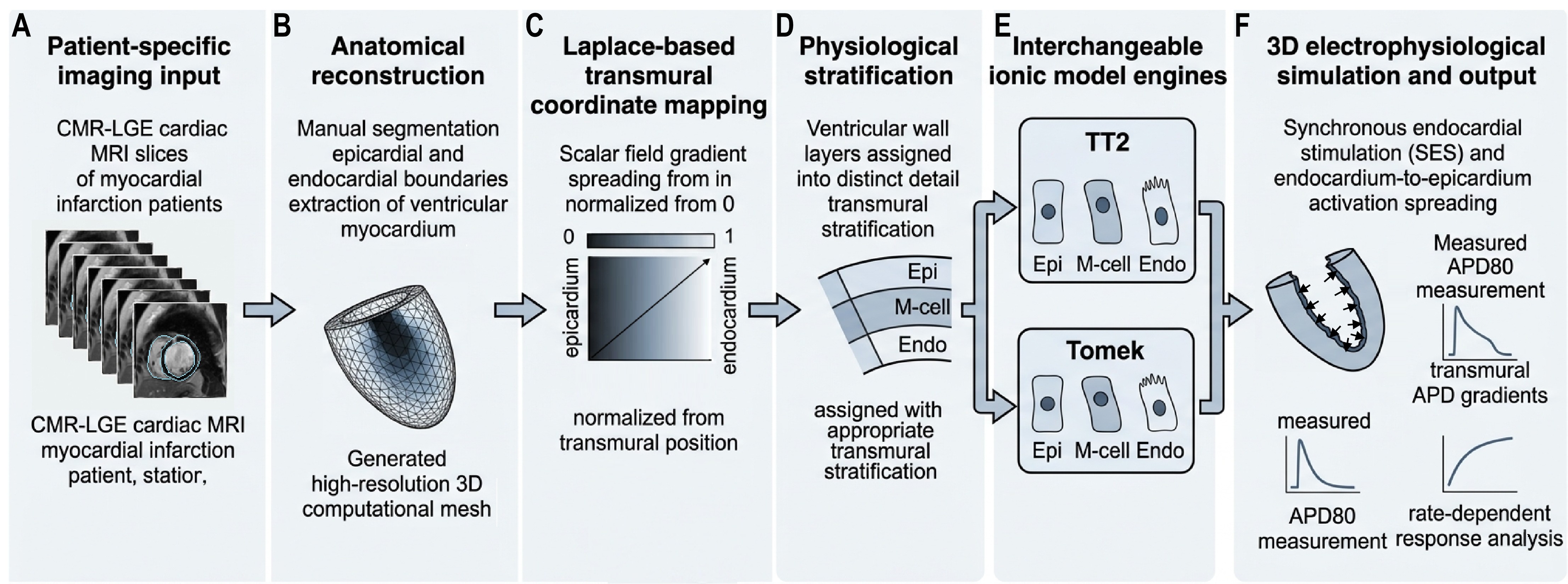}
\caption{Multi-scale patient-specific cardiac electrophysiological modeling workflow. The integrated pipeline consists of six consecutive functional panels: (A) Patient-specific imaging input utilizing clinical CMR-LGE cardiac MRI slices from myocardial infarction patients; (B) Anatomical reconstruction involving manual segmentation of epicardial and endocardial boundaries to generate a high-resolution 3D computational mesh; (C) Laplace-based transmural coordinate mapping to establish a normalized scalar field gradient from the epicardium (0) to the endocardium (1); (D) Physiological stratification where the ventricular wall is assigned into distinct transmural layers (Epi, M-cell, and Endo); (E) Integration of interchangeable ionic model engines deploying either the TT2 or Tomek cellular formulations; and (F) 3D electrophysiological simulation and output utilizing SES to evaluate transmural $\mathrm{APD}_{80}$ gradients and rate-dependent adaptation profiles.}
\label{fig:workflow}
\end{figure}

\subsection{Patient-Specific Three-Dimensional Ventricular Reconstruction}

The anatomical foundation of the CDT was established using patient-specific data to ensure high structural fidelity. High-resolution cardiac magnetic resonance-late gadolinium enhancement (CMR-LGE) images were acquired using a 3.0 T clinical scanner (Sonata, Siemens, Erlangen, Germany).

The reconstruction workflow involved the manual segmentation of the epicardial and endocardial boundaries by clinical experts, as shown in Figure~\ref{fig:workflow}A. The myocardial volume, defined as the region between these segmented boundaries, was then resampled to a high-resolution computational grid with a resolution of approximately 0.4 mm using the CardioViz3D platform(Figure~\ref{fig:workflow}B) ~\cite{toussaint2008integrated} (INRIA, Sophia Antipolis, France). The detailed methodology for this construction process has been published previously~\cite{deng2019characterizing,tong2021preliminary,miao2026theoretical}. Subsequent geometry processing and mesh refinement were performed using custom-developed software in MATLAB (Version 2021a, MathWorks Inc., Natick, MA, USA) to ensure the model was suitable for large-scale electrophysiological simulations.

\subsection{Laplace-Based Transmural Stratification}

To represent the electrophysiological heterogeneity of the myocardium, we employed a Laplace-Dirichlet-based approach to define a continuous scalar field within the ventricular wall~\cite{deng2019characterizing,tong2021preliminary,miao2026theoretical}. This field provides a normalized coordinate representing the relative transmural distance from the epicardium to the endocardium, regardless of local variations in wall thickness, as shown in Figure~\ref{fig:workflow}C. 

Based on a previously optimized transmural partition ratio established through a separate calibration study, the ventricular region was segmented into Epi, M-cell, and Endo domains (Figure~\ref{fig:workflow}D). This optimized ratio was held fixed across all TT2- and Tomek-based simulations because the present study focused on comparing ionic-model-dependent rate responses rather than repeating the stratification optimization procedure. This stratified architecture enables the CDT to simulate distinct repolarization patterns and transmural APD gradients under varying pacing conditions.
\subsection{Cellular Electrophysiological Models}

All simulations in this study were conducted using the open-source software platform openCARP~\cite{plank2021opencarp}. To accurately simulate the repolarization dynamics of human ventricular myocytes within a digital twin framework, two established human ventricular ionic models were employed as simulation engines: the TT2 model and the Tomek model. For both models, specific parameter sets corresponding to the Epi, Mcell, and Endo phenotypes were implemented to represent transmural cellular heterogeneity, as shown in Figure~\ref{fig:workflow}E. These two ionic engines exhibit substantial differences in the mathematical formulations of key ion currents, such as $I_{Kr}$ ,$I_{Ks}$ and $I_{CaL}$. These variations lead to distinct rate-dependent characteristics and APD adaptation patterns, which serve as a critical baseline for evaluating model-dependent outcomes across different scales.

\subsection{Stimulation Protocol}
To ensure ionic stability and eliminate transient artifacts, a rigorous pre-conditioning protocol was implemented. For the single-cell models, $V_{rest}$ and APD were measured from the 100th paced beat at each selected BCL, after the ionic state had reached a stable condition. The corresponding steady-state ionic variables were subsequently used to initialize the 3D ventricular simulations. Following initialization, five organ-level pacing stimuli were applied to reduce tissue-level initialization transients, and APD in the 3D simulations was measured from the sixth propagated action potential.

The 3D simulations were conducted using the reconstructed ventricular model with transmural phenotype assignments. To approximate the rapid ventricular activation mediated by the His--Purkinje system (HPS), we implemented a multi-point Simultaneous Endocardial Stimulation (SES) strategy, as illustrated in the upper-left panel of Figure~\ref{fig:workflow}F. Under physiological conditions, electrical excitation propagates rapidly through the HPS and enters the ventricular myocardium through discrete Purkinje--myocardial junctions (PMJs), resulting in near-synchronous activation of the endocardial surface~\cite{vergara2014,wang2025}.

To functionally reproduce this activation pattern without explicitly reconstructing the Purkinje network, all computational nodes located on the left ventricular endocardial surface were initially selected as candidate stimulation sites. A sparsification algorithm was subsequently applied to filter the candidate set, yielding a limited and spatially distributed set of initial stimulation sites. These retained sites were activated simultaneously at the beginning of each pacing cycle, generating multiple endocardial wavefronts that rapidly propagated through the ventricular myocardium. This strategy preserves the near-synchronous nature of HPS-mediated activation while approximating the discrete spatial distribution of PMJs. It therefore reduces the influence of nonphysiological focal activation and enables transmural APD gradients to be evaluated under a physiologically representative activation pattern.

To investigate rate-dependent effects on APD and transmural heterogeneity, simulations were performed at BCLs of 600 ms, 1000 ms, 2000 ms, and 4000 ms.

\subsection{APD Measurement and Transmural Sampling}

To quantify the repolarization characteristics within the multi-scale framework, APD was defined as the time interval from the maximum upstroke velocity ($dV/dt_{\max}$) to 80\% repolarization ($\mathrm{APD}_{80}$). 

For the single-cell simulations, APD values were directly extracted from the stable voltage-time curves of the Epi, Mcell, and Endo components. In the 3D cardiac simulation, a structured transmural sampling strategy was employed to assess regional heterogeneity. The left ventricular lateral wall was computationally divided into ten transmural layers, spanning from the epicardium to the endocardium. Within each layer, the mean APD was derived from a set of randomly selected points using a spatially uniform distribution to ensure representative sampling and eliminate selection bias.

Transmural repolarization heterogeneity was quantified using the transmural APD gradient ($\Delta\mathrm{APD}$), defined as the difference between the maximum and minimum layer-wise mean APD values across the ten layers ($\Delta\mathrm{APD} = \mathrm{APD}_{\max} - \mathrm{APD}_{\min}$). This multi-scale sampling protocol allows for a direct comparison of how the ionic model engines (TT2 versus Tomek) translate from cellular dynamics to 3D tissue-level gradients under varying rate-dependent conditions. 

The experimental canine-heart data reported by Kong et al.~\cite{kong2009transmural} were quantified using $\mathrm{APD}_{60}$, whereas the present simulations used $\mathrm{APD}_{80}$. Therefore, these datasets were compared only with respect to the direction and qualitative pattern of transmural APD changes across pacing cycle lengths. No direct comparison of absolute APD values or calculation of pointwise errors was performed.

\section{Results}

\subsection{Intrinsic Rate-Dependent Behavior at the Single-Cell Scale}
At the single-cell scale, the TT2 and Tomek models exhibit model-specific response profiles across varying BCL conditions, as summarized in Table~\ref{tab:singlecell_apd}. The resting membrane potential ($V_{rest}$) of the Tomek model consistently maintains a more hyperpolarized state (ranging from -88.34 mV to -88.92 mV) compared to the TT2 model (-84.90 mV to -86.32 mV).

\begin{table}[t]
\caption{Single-cell APD and $\Delta\mathrm{APD}$ values for TT2 and Tomek models under different BCL conditions.}
\label{tab:singlecell_apd}
\centering
\scriptsize
\setlength{\tabcolsep}{2.6pt}
\begin{tabular*}{\textwidth}{@{\extracolsep{\fill}}lrrrrrrr@{}}
\toprule
\multicolumn{8}{c}{(a) TT2 single-cell model under different BCL conditions}\\
\midrule
100 beats & \multicolumn{2}{c}{Endo} & \multicolumn{2}{c}{M-cell} & \multicolumn{2}{c}{Epi} & $\Delta\mathrm{APD}$\\
 & $V_{rest}$ & APD & $V_{rest}$ & APD & $V_{rest}$ & APD & \\
 & (mV) & (ms) & (mV) & (ms) & (mV) & (ms) & (ms)\\
\midrule
BCL500  & -84.93 & 285.67 & -84.75 & 356.55 & -84.90 & 286.28 & 70.88\\
BCL600  & -85.12 & 297.03 & -84.92 & 383.21 & -85.08 & 297.94 & 86.18\\
BCL1000 & -85.44 & 305.69 & -85.26 & 410.46 & -85.40 & 306.56 & 104.77\\
BCL2000 & -85.89 & 305.06 & -85.43 & 418.38 & -85.92 & 308.54 & 113.32\\
BCL3000 & -86.16 & 304.26 & -86.10 & 423.81 & -86.15 & 306.61 & 119.55\\
BCL4000 & -86.32 & 306.10 & -86.27 & 428.68 & -86.31 & 307.97 & 122.58\\
BCL5000 & -86.42 & 307.24 & -86.38 & 432.24 & -86.41 & 309.04 & 125.00\\
\midrule
\multicolumn{8}{c}{(b) Tomek single-cell model under different BCL conditions}\\
\midrule
100 beats & \multicolumn{2}{c}{Endo} & \multicolumn{2}{c}{M-cell} & \multicolumn{2}{c}{Epi} & $\Delta\mathrm{APD}$\\
 & $V_{rest}$ & APD & $V_{rest}$ & APD & $V_{rest}$ & APD & \\
 & (mV) & (ms) & (mV) & (ms) & (mV) & (ms) & (ms)\\
\midrule
BCL500  & -88.77 & 231.23 & -88.34 & 324.53 & -88.86 & 210.98 & 113.55\\
BCL600  & -88.82 & 243.27 & -88.43 & 333.18 & -88.92 & 219.34 & 113.84\\
BCL1000 & -88.80 & 271.64 & -88.54 & 348.04 & -88.89 & 245.22 & 102.82\\
BCL2000 & -88.63 & 300.77 & -88.52 & 380.34 & -88.73 & 285.72 & 94.62\\
BCL3000 & -88.52 & 317.91 & -88.46 & 397.40 & -88.64 & 305.15 & 92.25\\
BCL4000 & -88.43 & 329.57 & -88.41 & 408.64 & -88.57 & 316.09 & 92.55\\
BCL5000 & -88.35 & 337.92 & -88.53 & 417.76 & -88.50 & 328.84 & 88.92\\
\bottomrule
\end{tabular*}
\end{table}

Furthermore, these two mathematical frameworks demonstrate divergent APD adaptation trajectories across the tested BCL range. The Tomek model exhibits a highly synchronous APD prolongation across all cell phenotypes (Endo, Mcell, and Epi) as BCL increases. In contrast, the TT2 model displays a heterogeneous, cell-type-specific response. While the Mcell APD in the TT2 model increases monotonically with BCL, the Endo and Epi cells manifest a biphasic trend, where APD initially prolongs but subsequently reaches a plateau or slightly declines at very long BCLs.

The most fundamental departure between the two models lies in the evolution of $\Delta\mathrm{APD}$. As BCL extends, the Tomek model predicts a progressive convergence of cellular APDs, resulting in a reduction of $\Delta\mathrm{APD}$. Conversely, the TT2 model predicts a consistent amplification of transmural heterogeneity, highlighting fundamentally different kinetic descriptions of rate-dependency in the representation of repolarizing currents.

\subsection{Comparative Evaluation against Experimental Benchmarks}

\begin{figure*}
\centering
\figurefile{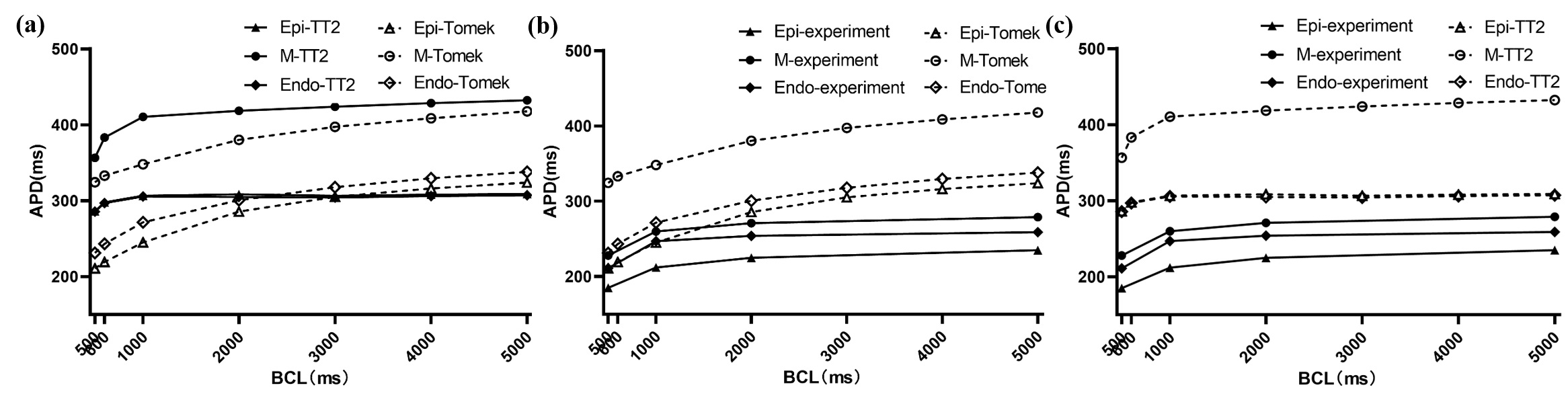}{Single-cell APD-BCL response and experimental validation panel}
\caption{Comparative analysis of APD rate-dependence between simulation models and experimental benchmarks. (a) Evolution of $\Delta\mathrm{APD}$ in TT2 and Tomek models under varying BCL conditions. (b) Assessment of the APD-BCL relationship for the Tomek single-cell model against clinical measurement data. Clinical benchmarks are derived from canine ventricular wedge preparations reported by Akar and Rosenbaum \cite{akar2003transmural}. (c) Corresponding comparison for the TT2 single-cell model against the same clinical dataset. In (b) and (c), experimental curves are reproduced based on the physiological trends observed in the source publication.}
\label{fig:singlecell_validation}
\end{figure*}

\vspace{-0.6em}

The comparative evaluation at the single-cell scale reveals that the TT2 and Tomek models offer distinct computational representations of ventricular rate-dependency when benchmarked against experimental data \cite{akar2003transmural}, as illustrated in Figure~\ref{fig:singlecell_validation}. The Tomek model successfully reproduces the physiological repolarization hierarchy (Mcell $>$ Endo $>$ Epi) throughout the standard physiological range, which aligns with established observations of transmural APD distribution. However, at extended BCLs (2000--5000 ms), the model manifests a sustained APD prolongation, indicating a divergence from the stable plateau behavior typically observed in experimental preparations.

In contrast, the TT2 model provides a different kinetic profile. While its biphasic adaptation in Epi and Endo cells presents a more complex response than the synchronous prolongation of the Tomek model, the TT2 framework more closely approximates the experimental plateau behavior at slow pacing rates. Additionally, the TT2 model reflects the amplification of transmural gradients under bradycardic conditions, capturing the expansion of repolarization dispersion across the myocardial wall. These results indicate that while the Tomek model captures the spatial ordering of repolarization, the TT2 model reflects the rate-dependent stabilization of APD and the associated intensification of transmural heterogeneity.

\subsection{Transmural Electrophysiological Responses to BCL in Three-Dimensional Models}

To characterize layer-specific APD and transmural heterogeneity within the stratified ventricular myocardium, the responses of the TT2 and Tomek models were evaluated using a 3D ventricular model at BCLs of 600, 2000, and 4000 ms.

\begin{figure}
\centering
\figurefile{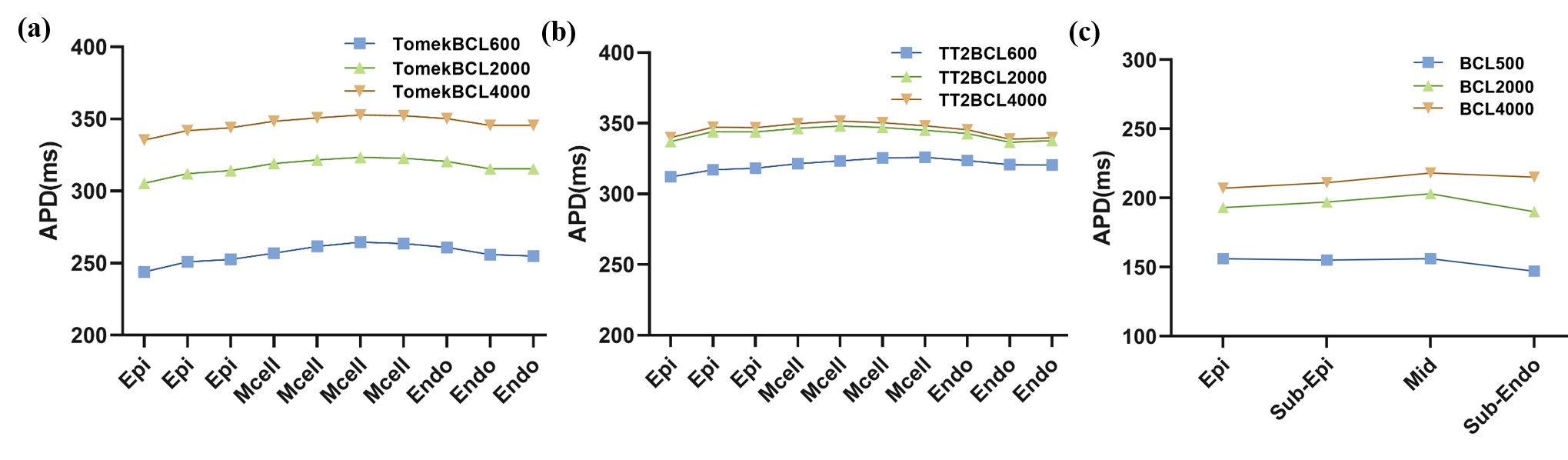}{3D transmural APD remodeling panel}
\caption{Tissue-scale APD remodeling in three-dimensional ventricles. (a) Tomek-based transmural $\mathrm{APD}_{80}$ profiles at representative BCLs. (b) TT2-based transmural $\mathrm{APD}_{80}$ profiles at representative BCLs. (c) Experimental transmural $\mathrm{APD}_{60}$ distributions in canine hearts at CLs of 500, 2000, and 4000 ms, approximately reconstructed from Kong et al.~\cite{kong2009transmural}. Because the experimental and computational APD definitions differ, panel (c) is used exclusively to compare the direction and spatial pattern of transmural changes rather than their absolute magnitudes.}
\label{fig:tissue}
\end{figure}

As illustrated in Figure~\ref{fig:tissue}a, the Tomek model exhibited a pronounced frequency dependence of APD across all myocardial layers, mirroring the trends observed in single-cell simulations. The mean inter-layer APD (defined as the mean $\pm$ standard deviation across ten layers) increased from $256.59 \pm 6.41$ ms at a BCL of 600 ms to $346.68 \pm 5.38$ ms at 4000 ms, representing a 35.11\% elevation. While the directional trend of $\Delta\mathrm{APD}$ remained consistent with single-cell results, the magnitude of the gradient was significantly modulated by intercellular coupling. Specifically, $\Delta\mathrm{APD}$ decreased from 20.67 ms to 17.26 ms as BCL prolonged, a narrower shift compared to the 21.29 ms reduction predicted at the single-cell scale. This discrepancy underscores the role of electrotonic coupling in attenuating frequency-dependent heterogeneity when transitioning across organizational scales.

Figure~\ref{fig:tissue}b demonstrates that the TT2 model manifested lower frequency dependence compared to the Tomek framework. As BCL increased from 600 to 4000 ms, the mean inter-layer APD rose from $320.84 \pm 4.20$ ms to $345.82 \pm 4.75$ ms, a 7.79\% increase. Notably, the 3D tissue environment altered the rate-dependent behavior of the Endo and Epi layers compared to their single-cell counterparts. Due to electrotonic interactions, these layers were influenced by the intrinsic kinetics of the Mcell layer; consequently, the mean APD of the Endo layer increased from $321.60 \pm 1.70$ ms to $341.27 \pm 3.62$ ms, while the Epi layer increased from $315.81 \pm 3.30$ ms to $344.74 \pm 4.16$ ms. These increments are substantially larger than the $\sim$10 ms elevations observed in isolated single cells. The resulting $\Delta\mathrm{APD}$ remained relatively stable, shifting marginally from 13.80 ms to 12.86 ms across the tested range.

\section{Discussion}

This study examines the multi-scale, rate-dependent electrophysiological behavior of the TT2 and Tomek human ventricular models, showing how single-cell kinetics are integrated under 3D syncytial electrotonic coupling. When moving from single-cell parameters to 3D stratified tissue, the results show that the two models provide different but complementary descriptions of rate-dependent adaptation. At the single-cell scale, the different APD adaptation patterns and the resulting $\Delta\mathrm{APD}$ are mainly caused by differences in $I_{Kr}$, $I_{Ks}$, and $I_{CaL}$ kinetics. The synchronous, monotonic APD prolongation of the Tomek model reflects a model designed to capture time-sensitive biomarkers and frequency-dependent tension dynamics \cite{coppini2013late}. In contrast, the cell-type-specific, biphasic pattern of the TT2 model keeps strong sensitivity to separate transmural phenotype parameters \cite{ten2004model}. These model differences affect the net repolarizing current balance at extended BCLs, leading to different transmural heterogeneity trends before tissue integration.

In the 3D stratified tissue, intercellular electrotonic coupling changes the behavior of each model in a scale-dependent way. Syncytial coupling in the 3D Tomek tissue acts as an electrotonic buffer, reducing single-cell $\Delta\mathrm{APD}$ compression while maintaining a smooth anatomical repolarization hierarchy. In the 3D TT2 model, electrotonic loading from the rate-responsive M-cell layer affects the adjacent Endo and Epi layers, which are less sensitive in isolated single-cell simulations. This interaction creates a tissue-level rate response that is not seen in isolation, resulting in clustered profiles at long BCLs. These changes show that 3D tissue structure can alter cellular rate-dependency and should be considered when using ionic models to predict arrhythmogenic substrates \cite{antzelevitch2004electrophysiological}.

\begin{credits}
\subsubsection{\ackname}
This  work  was  supported  by the Fundamental Research Funds for the Central Universities (DUT25YG231  to  D.  Deng),  the”Leading  Goose”  R\&D Program of Zhejiang Province (2025C02150 to D. Deng), the National Natural Science Foundation of China (62171408  to L. Xia), the  Key  Research  and  Development Program  of Zhejiang  Province  (2020C03016,  2023C03088, and  2025C02150  to  L.  Xia).

\subsubsection{\discintname}
The authors have no competing interests to declare that are relevant to the
content of this article.
\end{credits}

\bibliographystyle{splncs04}
\bibliography{references}

@article{ten2006alternans,
  author  = {ten Tusscher, Kirsten H. W. J. and Panfilov, Alexander V.},
  title   = {Alternans and spiral breakup in a human ventricular tissue model},
  journal = {American Journal of Physiology-Heart and Circulatory Physiology},
  volume  = {291},
  number  = {3},
  pages   = {H1088--H1100},
  year    = {2006},
  doi     = {10.1152/ajpheart.00109.2006}
}

@article{tomek2019development,
  author  = {Tomek, Jakub and Bueno-Orovio, Alfonso and Passini, Elisa and others},
  title   = {Development, calibration, and validation of a novel human ventricular myocyte model in health, disease, and drug block},
  journal = {eLife},
  volume  = {8},
  pages   = {e48890},
  year    = {2019},
  doi     = {10.7554/eLife.48890}
}

@article{akar2003transmural,
  author    = {Akar, Fadi G. and Rosenbaum, David S.},
  title     = {Transmural electrophysiological heterogeneities underlying arrhythmogenesis in heart failure},
  journal   = {Circulation Research},
  volume    = {93},
  number    = {7},
  pages     = {638--645},
  year      = {2003},
  publisher = {Lippincott Williams \& Wilkins},
  doi       = {10.1161/01.RES.0000092248.59479.AE}
}

@article{maharaj2008role,
  author  = {Maharaj, Thushka and Blake, Robert and Trayanova, Natalia A. and Gavaghan, David and Rodriguez, Blanca},
  title   = {The role of transmural ventricular heterogeneities in cardiac vulnerability to electric shocks},
  journal = {Progress in Biophysics and Molecular Biology},
  volume  = {96},
  number  = {1-3},
  pages   = {321--338},
  year    = {2008},
  doi     = {10.1016/j.pbiomolbio.2007.07.017}
}

@article{kong2009transmural,
  title={Transmural optical measurements of Vm dynamics during long-duration ventricular fibrillation in canine hearts},
  author={Kong, Wei and Ideker, Raymond E and Fast, Vladimir G},
  journal={Heart rhythm},
  volume={6},
  number={6},
  pages={796--802},
  year={2009},
  publisher={Elsevier},
  doi={https://doi.org/10.1016/j.hrthm.2009.02.028}
}

@article{deng2019characterizing,
  title={Characterizing conduction channels in postinfarction patients using a personalized virtual heart},
  author={Deng, Dongdong and Prakosa, Adityo and Shade, Julie and Nikolov, Plamen and Trayanova, Natalia A},
  journal={Biophysical Journal},
  volume={117},
  number={12},
  pages={2287--2294},
  year={2019},
  publisher={Elsevier},
  doi={10.1016/j.bpj.2019.07.024.}
}

@article{tong2021preliminary,
  title={Preliminary Study: Learning the Impact of Simulation Time on Reentry Location and Morphology Induced by Personalized Cardiac Modeling},
  author={Tong, Lv and Zhao, Caiming and Fu, Zhenyin and Dong, Ruiqing and Wu, Zhenghong and Wang, Zefeng and Zhang, Nan and Wang, Xinlu and Cao, Boyang and Sun, Yutong and others},
  journal={Frontiers in Physiology},
  volume={12},
  pages={733500},
  year={2021},
  publisher={Frontiers Media SA},
  doi={10.3389/fphys.2021.733500.}
}

@article{antzelevitch2004electrophysiological,
  title={Electrophysiological effects of ranolazine, a novel antianginal agent with antiarrhythmic properties},
  author={Antzelevitch, Charles and Belardinelli, Luiz and Zygmunt, Andrew C and Burashnikov, Alexander and Di Diego, Jos{\'e} M and Fish, Jeffrey M and Cordeiro, Jonathan M and Thomas, George},
  journal={Circulation},
  volume={110},
  number={8},
  pages={904--910},
  year={2004},
  publisher={Lippincott Williams \& Wilkins},
  doi={https://doi.org/10.1161/01.CIR.0000139333.83620.5D}
}

@article{glukhov2010transmural,
  title={Transmural dispersion of repolarization in failing and nonfailing human ventricle},
  author={Glukhov, Alexey V and Fedorov, Vadim V and Lou, Qing and Ravikumar, Vinod K and Kalish, Paul W and Schuessler, Richard B and Moazami, Nader and Efimov, Igor R},
  journal={Circulation research},
  volume={106},
  number={5},
  pages={981--991},
  year={2010},
  publisher={Lippincott Williams \& Wilkins},
  doi={https://doi.org/10.1161/CIRCRESAHA.109.204891}
}

@article{ten2004model,
  title={A model for human ventricular tissue},
  author={Ten Tusscher, Kirsten HWJ and Noble, Denis and Noble, Peter-John and Panfilov, Alexander V},
  journal={American Journal of Physiology-Heart and Circulatory Physiology},
  volume={286},
  number={4},
  pages={H1573--H1589},
  year={2004},
  publisher={American Physiological Society},
  doi={https://doi.org/10.1152/ajpheart.00794.2003}
}

@article{coppini2013late,
  title={Late sodium current inhibition reverses electromechanical dysfunction in human hypertrophic cardiomyopathy},
  author={Coppini, Raffaele and Ferrantini, Cecilia and Yao, Lina and Fan, Peidong and Del Lungo, Martina and Stillitano, Francesca and Sartiani, Laura and Tosi, Benedetta and Suffredini, Silvia and Tesi, Chiara and others},
  journal={Circulation},
  volume={127},
  number={5},
  pages={575--584},
  year={2013},
  publisher={Lippincott Williams \& Wilkins Hagerstown, MD},
  doi={https://doi.org/10.1161/CIRCULATIONAHA.112.134932}
}

@article{miao2026theoretical,
  author  = {Miao, Y. and Fu, Z. and Zhang, J. and others},
  title   = {Theoretical quantitative model and clinical outcome predictions of conductive cardiac patches for electrophysiological treatments},
  journal = {Nature Biomedical Engineering},
  year    = {2026},
  doi     = {10.1038/s41551-026-01659-x}
}

@inproceedings{toussaint2008integrated,
  title={An Integrated Platform for Dynamic Cardiac Simulation and Image Processing: Application to Personalised Tetralogy of Fallot Simulation.},
  author={Toussaint, Nicolas and Mansi, Tommaso and Delingette, Herv{\'e} and Ayache, Nicholas and Sermesant, Maxime},
  booktitle={VCBM},
  pages={21--28},
  year={2008},
  doi={https:///doi.org/10.2312/VCBM/VCBM08/021-028}
}

@article{plank2021opencarp,
  title={The openCARP simulation environment for cardiac electrophysiology},
  author={Plank, Gernot and Loewe, Axel and Neic, Aurel and Augustin, Christoph and Huang, Yung-Lin and Gsell, Matthias AF and Karabelas, Elias and Nothstein, Mark and Prassl, Anton J and S{\'a}nchez, Jorge and others},
  journal={Computer methods and Programs in Biomedicine},
  volume={208},
  pages={106223},
  year={2021},
  publisher={Elsevier},
  doi={https://doi.org/10.1016/j.cmpb.2021.106223}
}

@article{vergara2014,
  author={Vergara, C. and Palamara, S. and Catanzariti, D. and Nobile, F. and Faggiano, E. and Pangrazzi, C. and Centonze, M. and Maines, M. and Quarteroni, A. and Vergara, G.},
  title={Patient-specific generation of the {Purkinje} network driven by clinical measurements of a normal propagation},
  journal={Medical \& Biological Engineering \& Computing},
  volume={52},
  number={10},
  pages={813--826},
  year={2014},
  month=oct,
  doi={10.1007/s11517-014-1183-5}
}

@article{wang2025,
  author={Wang, Zhinuo Jenny and Holmes, Maxx and Doste, Ruben and Camps, Julia and Margara, Francesca and Vazquez, Mariano and Rodriguez, Blanca},
  title={Calibration and validation strategy for electromechanical cardiac digital twins},
  journal={eLife},
  volume={14},
  pages={RP106555},
  year={2025},
  doi={10.7554/eLife.106555.2}
}

\end{document}